\documentclass[acmsmall,nonacm]{acmart}
\usepackage{algorithm}      
\usepackage{algpseudocode}  
\usepackage{enumitem}
\usepackage{graphicx}
\usepackage{booktabs}
\usepackage{multirow}
\usepackage{subcaption}
\usepackage{caption}
\usepackage{geometry}
\usepackage{tikz}
\usepackage{pgfplots}
\pgfplotsset{compat=1.17}
\usetikzlibrary{patterns}
\newcommand{\mycomment}[1]{}
\usepackage{xcolor}     
\usepackage{cprotect}   
\newcommand{\codebox}[1]{%
  \colorbox{gray!15}{\ttfamily\small\sffamily #1}%
}
\AtBeginDocument{%
  }

\newcounter{row}
\newcounter{col}

\algnewcommand{\LeftComment}[1]{\Statex \(\triangleright\) #1}

\begin{document}

\title{Efficient Column-Wise N:M Pruning on RISC-V CPU}
\footnote{New Paper, Not an Extension of a Conference Paper}

\author{Chi-Wei Chu}
 \affiliation{%
   \institution{Institute of Information Science, Academia Sinica}
   \country{Taiwan}
}
\email{wewe@iis.sinica.edu.tw}
\author{Ding-Yong Hong}
 \affiliation{%
   \institution{Institute of Information Science, Academia Sinica}
   \country{Taiwan}
}
\email{dyhong@iis.sinica.edu.tw}
\author{Jan-Jan Wu}
 \affiliation{%
   \institution{Institute of Information Science, Academia Sinica}
   \country{Taiwan}
}
\email{wuj@iis.sinica.edu.tw}

\begin{abstract}
In deep learning frameworks, weight pruning is a widely used technique for improving computational efficiency by reducing the size of large models.
This is especially critical for convolutional operators, which often act as performance bottlenecks in convolutional neural networks (CNNs). 
However, the effectiveness of pruning heavily depends on how it is implemented, as different methods can significantly impact both computational performance and memory footprint.
In this work, we propose a column-wise N:M pruning strategy applied at the tile level and modify XNNPACK to enable efficient execution of pruned models on the RISC-V vector architecture.
Additionally, we propose fusing the operations of im2col and data packing to minimize redundant memory accesses and memory overhead.
To further optimize performance, we incorporate AITemplate's profiling technique to identify the optimal implementation for each convolutional operator.
Our proposed approach effectively increases ResNet inference throughput by as much as $4.0\times$, and preserves ImageNet top-1 accuracy within 2.1\% of the dense baseline.
\end{abstract}





\received{19 March 2025}

\maketitle
\section{Introduction}\label{sec:introduction}

The rapid advancement in deep learning has led to a substantial increase in neural network size and capability, resulting in growing demands for computing power and memory resources~\cite{han2015learning, gale2019state}.
This growing complexity presents significant challenges for model deployment, especially on platforms with limited computational and memory capacities.
Consequently, neural network compression has become essential to enable deployment on resource-constrained devices.

Weight pruning is a widely used technique that introduces sparsity into neural networks by selectively removing redundant or less significant weights.
By eliminating model parameters, pruning reduces both computational overhead and memory footprint, thus leading to improved execution performance.
This optimization is particularly beneficial for real-time applications, where computational efficiency and low latency are critical requirements.

Weight pruning can be broadly classified into two categories: {\em unstructured pruning} and {\em structured pruning}.
Unstructured pruning removes individual weights without a specific pattern, while structured pruning applies predefined sparsity structures to optimize execution on modern accelerators.
Among structured pruning techniques, NVIDIA's 2:4 pruning is a prominent method, which enforces that at least two out of every four consecutive weights in a layer are pruned, maintaining a fine-grained structured format.
To support this format, NVIDIA introduced the Sparse Tensor Core~\cite{nvidia2020structured}, a specialized hardware unit designed to efficiently process models with 2:4 sparsity.
Building upon this foundation, the 2:4 pruning has been generalized to an N:M pruning scheme, which offers improved flexibility~\cite{sun2021domino, zhou2021nm, hubara2021transposable, chmiel2023mvu, huang2024elsa}.
Recent studies have also shown that N:M pruning can be integrated with other compression schemes, such as block-based and vector-based formats, to achieve even higher compression ratios while remaining executable on the GPUs even without native N:M support~\cite{castro2022venom,chen2025accelerating}.

Despite its success in GPU-based acceleration, the effectiveness of N:M pruning for CPU architectures with SIMD (single-instruction multiple-data) capabilities remains underexplored.
Hence, this work aims to bridge this gap by designing and evaluating an efficient N:M pruning approach optimized for CPUs.
We adopt CPUs because they remain ubiquitous and essential processing units across a wide range of platforms, from edge devices to enterprise servers.
Additionally, their general-purpose computing capabilities make them a suitable target for enabling N:M sparse model execution, without the need for specialized accelerators.
Unlike previous works~\cite{castro2022venom, chen2025accelerating} that rely on Sparse Tensor Cores and are limited to specific N:M sparsity patterns, our CPU-based approach offers greater flexibility, supporting a wider range of N:M configurations.
Furthermore, most of prior works mainly focuses on improving N:M pruning accuracy through training techniques~\cite{sun2021domino, bambhaniya2024progressivegradientflowrobust, lu2023steplearningnmstructured, kao2022trainingrecipenmstructured}.
In contrast, our study aims to improve execution efficiency, and thus complements these prior efforts. 

In this work, we target the RISC-V architecture and its Vector extension, because RISC-V is gaining increasing attention for its efficiency and scalability.
While RISC-V has been widely adopted in embedded systems due to its energy efficiency, it is increasingly being deployed in high-end server platforms.
In addition, this work focuses on accelerating convolutional neural networks (CNNs) with N:M pruning, which are particularly impacted by the high computational cost of convolution operations.

Designing N:M pruning for RISC-V architectures should consider the following critical issues. First, embedded systems based on RISC-V are highly sensitive to memory access overheads~\cite{conti2018xne, flamand2018gap}.
Thus, it is crucial to minimize the memory overhead introduced by pruning so as to fully realize the performance gain.
Second, N:M pruning inherently involves indirect data accesses, which can potentially lead to redundant memory loads.
Third, the NHWC tensor layout, which is optimized for dense operations on CPUs, is not well-suited for sparse tensor computations due to non-contiguous memory accesses.
Finally, the RISC-V Vector architecture introduces unique features, such as scalable (i.e., sizeless) vector processing and register grouping (LMUL).
Effectively exploiting the RISC-V Vector necessitates optimized strategies to enhance computational efficiency and register usage.

To address these challenges, we propose \textbf{column-wise N:M pruning}, a software-hardware co-optimization strategy to enable efficient execution of N:M sparse models on RISC-V CPUs.
To mitigate the problem of redundant memory loads, we extend traditional row-based N:M pruning to a {\em column-wise} pruning strategy, where elements within columns are grouped and subsequently pruned or retained as a unit. 
By maintaining such regular patterns within columns, our approach maximizes data reuses and avoids data reloading.
Moreover, our approach supports significantly larger pruning groups (M), which can span the entire input channel dimensions, thus making it more akin to unstructured pruning while retaining structured execution advantages.
To further reduce memory overhead, we introduce a fusion strategy that combines the two memory-intensive operations, {\em im2col} and {\em data packing}, into a single memory-efficient step.
Finally, we design our optimizations in a framework that integrates XNNPACK~\cite{xnnpack_github} and the AITemplate~\cite{aitemplate_github} AI compiler.

In summary, this paper makes the following contributions:
\begin{itemize}[itemsep=0pt,nosep]
\item We propose a novel column-wise N:M pruning method, which enhances data reuses and minimizes redundant memory accesses. Our method supports arbitrary N:M sparsity patterns and improves model accuracy.
\item We propose to combine the im2col and data packing operations into one step. Our method significantly reduces memory overhead.
\item We design and evaluate our approach on the AITemplate AI compiler, extended with the XNNPACK backend to support execution on RISC-V. Furthermore, we leverage AITemplate’s tuning mechanism to determine the optimal parameters for RISC-V, including tile sizes and vector register group multipliers (LMUL).
\item Experimental results indicate that our approach reduces L1-cache loads by up to $42\%$, achieves speedups of up to $4\times$, and maintains an accuracy loss of $\le 2.1\%$ for sparse ResNet models on the ImageNet dataset compared to the dense counterpart.
\end{itemize}

The remainder of this paper is organized as follows.
Section~\ref{sec:related} reviews background and related work.
Section~\ref{sec:method} introduces our proposed column-wise N:M pruning method and fusion optimization, and describes the implementation details.
Section~\ref{sec:experiment} evaluates our method.
Section \ref{sec:discussion} discusses the impact of data layouts.
Section~\ref{sec:conclusion} concludes the paper.

\section{Background and Related Works}\label{sec:related}

This section reviews the background and related works, including weight pruning, GEMM-based convolution, packing methods, and the RISC-V Vector extension.

\subsection{Weight Pruning}

\begin{figure}[t!]
    \begin{minipage}[b]{0.5\linewidth}
    \centering
    \includegraphics[width=\textwidth]{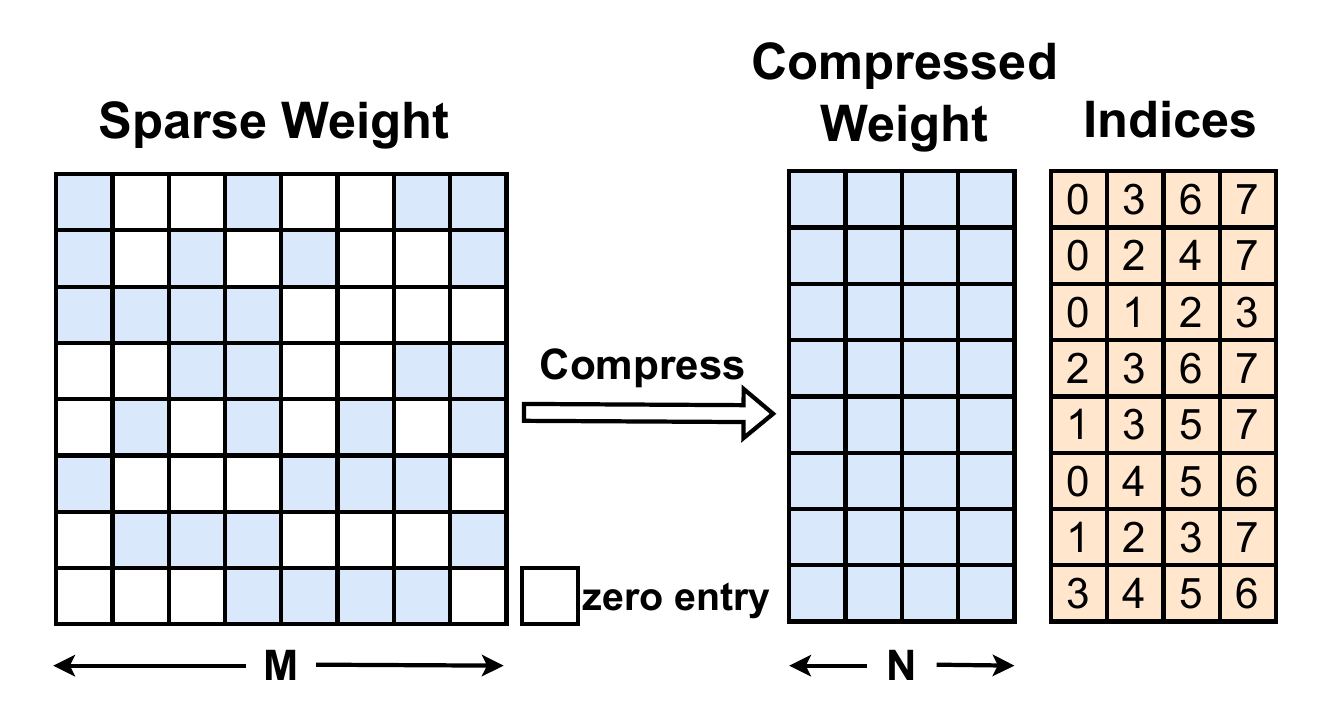}
    \caption{N:M pruning with N=4 and M=8.}
    \label{fig:Compressed_Weight}
    \Description{Diagram illustrating sparse weight being compressed.}
    \end{minipage}
    \begin{minipage}[b]{0.46\linewidth}
      \centering
      \includegraphics[width=.96\textwidth]{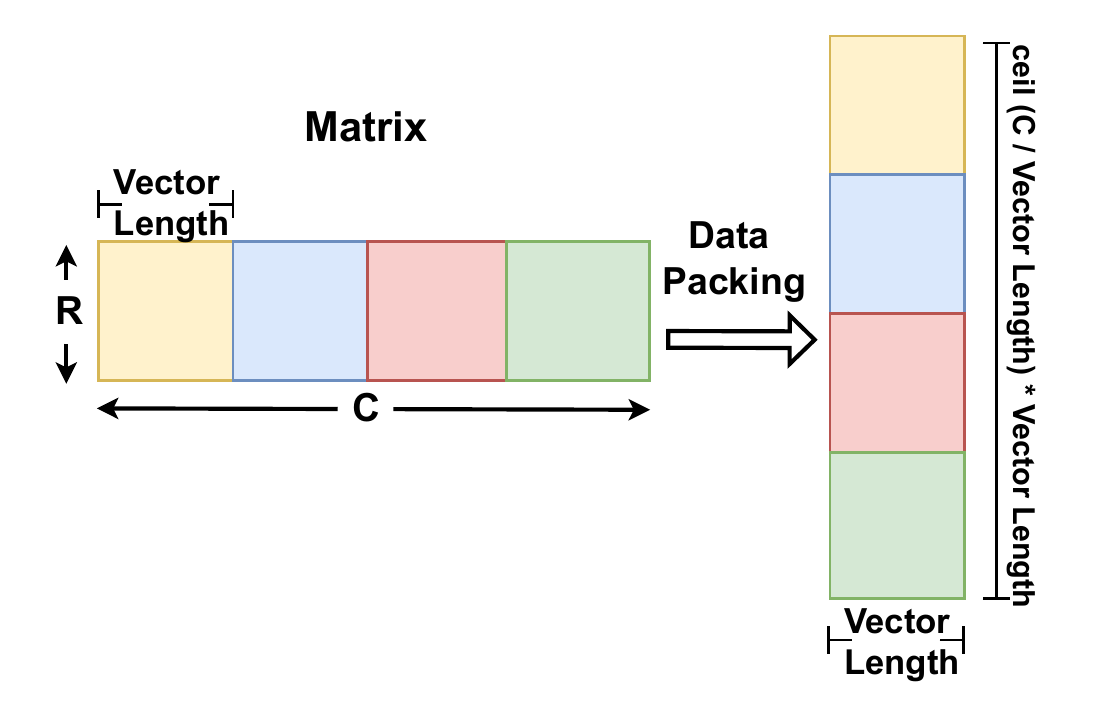}
      \caption{Data packing based on vector length.}
      \label{fig:Data_Packing}
      \Description{Diagram illustrating data packing for aligning matrix data with SIMD vector length.}
    \end{minipage}
\end{figure}

\mycomment{
\begin{figure}[t!]
\centering
  \begin{subfigure}[t]{0.48\textwidth}
    \centering
    \includegraphics[width=\textwidth]{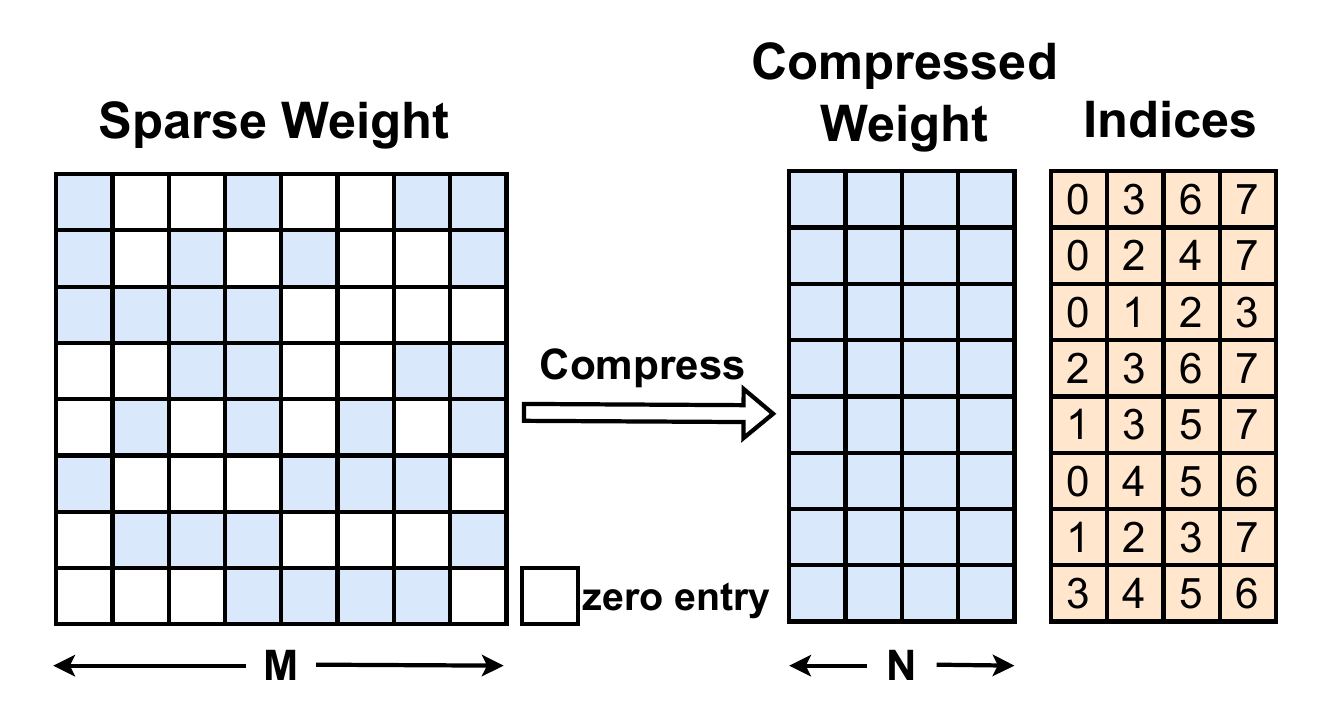}
    \caption{Compressed Weight and indices}
    \label{fig:Compressed_Weight}
    \Description{Diagram illustrating sparse weight being compressed.}
  \end{subfigure}
  \hfill
  \begin{subfigure}[t]{0.48\textwidth}
      \centering
      \includegraphics[width=\textwidth]{figure/Data_packing.pdf}
      \vspace{-3mm}
      \caption{Illustration of  Data packing conforming to the SIMD instruction vector length.}
      \label{fig:Data_Packing}
      \Description{Diagram illustrating data packing for aligning matrix data with SIMD vector length.}
  \end{subfigure}
  \caption{}
  \vspace{-5mm}
\end{figure}
}

Weight pruning techniques can be classified into unstructured and structured pruning, based on the criteria used to remove neural network parameters.

{\bf Unstructured pruning} eliminates individual weights without enforcing a specific pattern.
This method can obtain better model accuracy due to its flexibility in sparsity distribution, and often aligns well with common sparse storage formats such as CSR (compressed sparse row) and CSC (compressed sparse column)~\cite{saad2003iterative}.
Prior studies have explored various metrics for assessing weight importance, including magnitude-based~\cite{han2015learning}, gradient-based ~\cite{molchanov2017pruningconvolutionalneuralnetworks, lee2019snipsingleshotnetworkpruning, wang2020pickingwinningticketstraining, tanaka2020synflow}, and hybrid methods that combine both ~\cite{sanh2020movement, belay2022gradmag, zhu2024fggp, guo2016dns}.
However, the lack of structured regularity often results in irregular memory access patterns, making hardware acceleration less effective~\cite{gale2019state, mishra2021accelerating}.
In contrast, this work focuses on the N:M structured pruning.
We demonstrate that our approach is able to achieve substantial improvements in both performance and hardware efficiency.

{\bf Structured pruning} removes weights according to predefined patterns, helping to mitigate the hardware inefficiencies associated with unstructured pruning.
Various structured pruning techniques have been proposed based on different structural granularities, such as layer-wise~\cite{dong2017layerwiseOBS}, channel-wise~\cite{he2017channelpruning}, and group-wise~\cite{mao2017exploring, he2018soft, liu2021group}.
Recently, N:M pruning has gained significant attention, where at most N elements are retained  within each group of M consecutive weights.
Figure~\ref{fig:Compressed_Weight} shows an example of 50\% sparsity with N=4 and M=8.
After pruning the model, the sparse weight is compressed into a compressed weight format and an index array.
To improve the accuracy of N:M pruned models, researchers have explored advanced training techniques~\cite{bambhaniya2024progressivegradientflowrobust, lu2023steplearningnmstructured, kao2022trainingrecipenmstructured} and non-uniform sparsity approaches~\cite{sun2021domino}.

\citet{castro2022venom} introduced VENOM, a method for executing N:M sparse models on Sparse Tensor Cores, which natively support only 2:4 sparsity.
VENOM employs a two-stage sparsification process: it first compresses the dense model into a smaller dense model with an arbitrary sparsity ratio, and then prunes the model again using the standard 2:4 pruning method~\cite{nvidia2020structured}.
However, this two-stage approach lacks the flexibility to represent arbitrary N:M sparsity patterns; for instance, it cannot support 3:4 sparsity.
In contrast, our method enables direct support for arbitrary N:M configurations.

For CPU architectures, \citet{titopoulos2023optimizing} extended RISC-V with custom instructions to facilitate N:M sparse model execution.
In contrast, our method is designed to run efficiently on existing RISC-V CPUs, without requiring any specialized hardware or instruction set modifications.
On the other hand, \citet{elsen2020fast} developed an optimized sparse matrix-dense matrix multiplication (SpMM) kernel using the Arm Neon SIMD extension.
Their approach prunes model weights along the output channel dimension using a block-based format, which enables data reuse of fetched input data.
Inspired by this method, we extend N:M pruning with a {\bf column-wise pruning format}, enhancing the computational efficiency of sparse models by aligning sparsity with memory access patterns. 

\subsection{GEMM-based Convolution and Data Packing}\label{subsec:related_gemm_packing}

In this work, we build our method by extending Google's XNNPACK framework~\cite{xnnpack_github}, which offers highly optimized deep learning operator implementations for CPUs, including x86, ARM, and RISC-V.
In this section, we describe two key optimizations used in XNNPACK: {\em GEMM-based convolution} and {\em data packing}.

GEMM-based convolution is a widely adopted technique for accelerating convolution operators by expressing them as general matrix multiplication (GEMM) operations.
At the core of this transformation is {\em im2col} (image-to-column), which extracts local receptive fields (i.e., patches) from input feature maps and rearranges them into columns of a patch matrix, while convolution kernels are flattened into rows of a filter matrix.
This restructuring enables the convolution to be efficiently executed as a matrix multiplication using highly optimized GEMM routines. 

However, constructing the patch matrix introduces non-trivial memory overhead, especially for kernels larger than 1×1~\cite{anderson2017low}.
To alleviate this issue, XNNPACK employs an alternative approach known as Indirect Convolution~\cite{dukhan2019indirectconvolutionalgorithm}, which utilizes an {\em indirection buffer} to store pointers to the relevant input channels in the feature maps.
Matrix multiplication is then performed by reading data directly from input feature maps via the indirection buffer, thus eliminating the need to explicitly construct the patch matrix and reducing the data rearrangement overhead.

Data packing is an optimization technique that reorganizes data into hardware-efficient layouts to improve cache locality, maximize memory bandwidth utilization, and facilitate parallel execution.
Packing methods typically reorder data into blocked, tiled, or vector-aligned formats before  computation.
In XNNPACK, matrices are repacked into vector-aligned layouts according to the SIMD vector length prior to GEMM operations, as illustrated in Figure~\ref{fig:Data_Packing}.

\subsection{RISC-V Vector Extension}\label{sec:rvv}

RISC-V Vector (RVV)~\cite{riscv_vector_v1_0} is an extension to the RISC-V instruction-set architecture.
It features vector-length agnostic (VLA) execution, which allows the same binary to run efficiently across hardware platforms of varying vector lengths, without the need of source code recompilation.
In addition, RVV supports the mechanism known as the vector register grouping multiplier (LMUL).
This feature allows multiple vector registers to be combined into a single logical register, enabling operations on wider data types and increasing processing throughput. 
For example, consider a RISC-V system with 32 vector registers, each 256 bits wide. 
Setting LMUL = 8 allows a single instruction to operate on 256×8 = 2048 bits of data. However, this grouping reduces the effective number of available vector registers to 32 / 8 = 4. 

\section{Methodology}\label{sec:method}

In this section, we first analyze the inefficiency of traditional N:M pruning for CPUs.
Then, we introduce our column-wise N:M pruning method to enhance execution efficiency.
Subsequently, we present the fusion method to reduce memory overhead.
Finally, we provide the implementation details of our auto-tuning framework.

\subsection{Column-Wise N:M Pruning}\label{subsec:method_Microkernel_Design}

\begin{figure*}[t!]
  \centering
  \begin{subfigure}[t]{\textwidth}
    \centering
    \includegraphics[width=\textwidth]{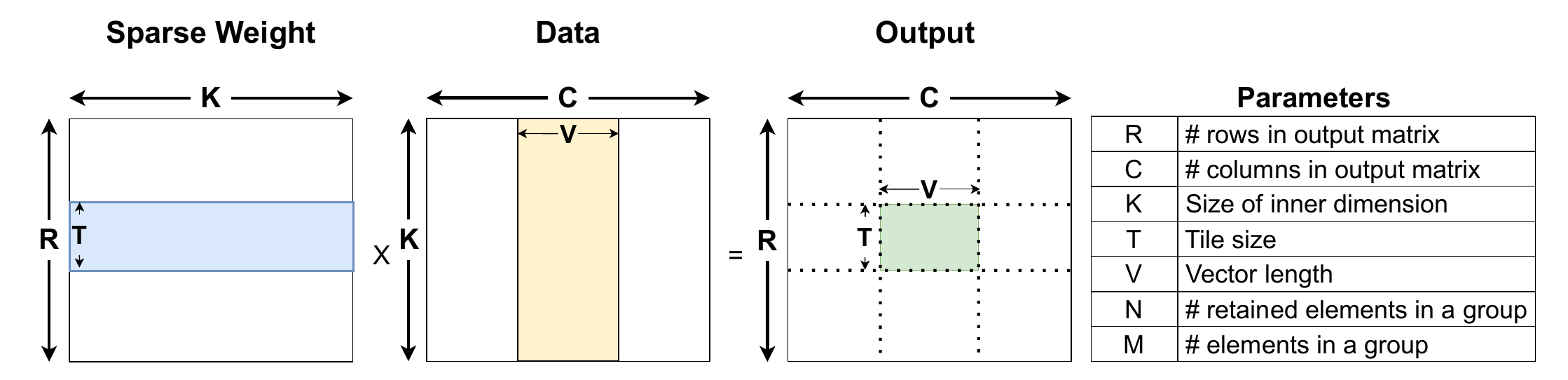}
    \caption{Matrix multiplication of sparse weight and dense data.}
    \label{fig:GEMM}
    \Description{GEMM}
  \end{subfigure}\\
  \begin{subfigure}[t]{.45\textwidth}
    \centering
    \includegraphics[width=\textwidth]{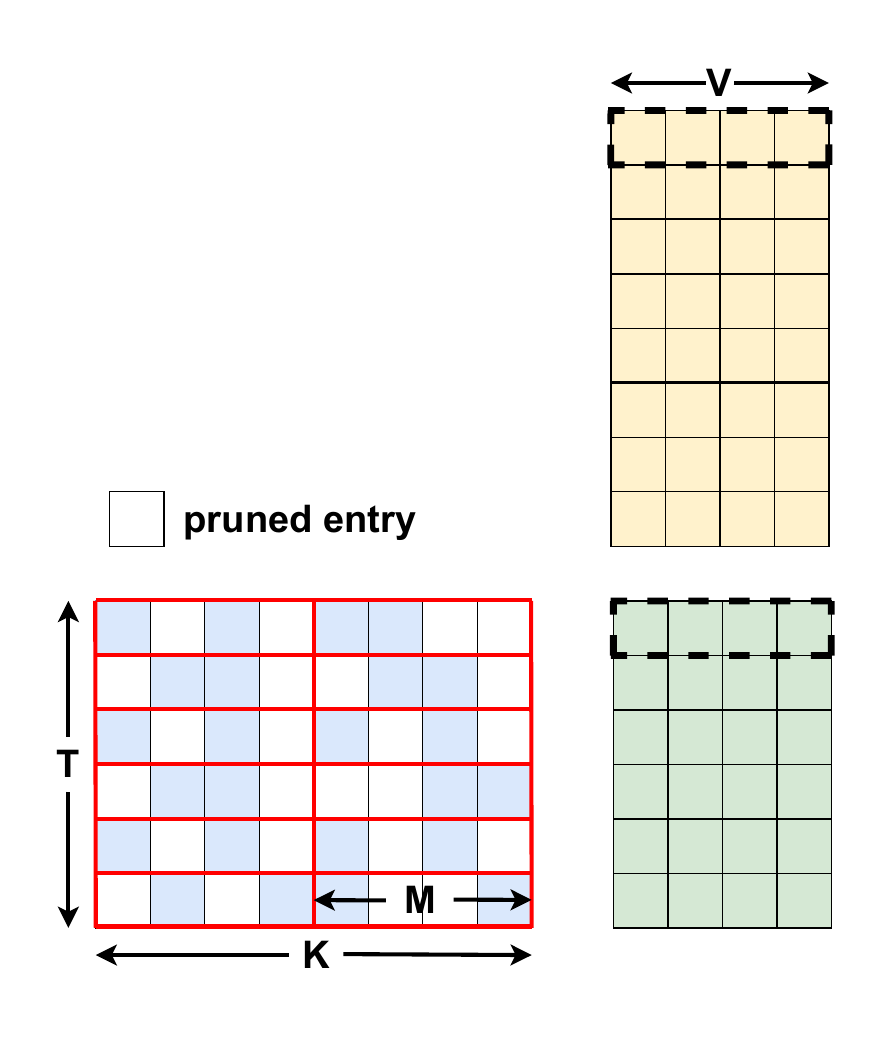}
    \caption{Row-based N:M pruning.\\(N=2 and M=4)}
    \label{fig:NMPruning}
    \Description{NMPruning}
  \end{subfigure}\hspace{2mm}
  \begin{subfigure}[t]{0.51\textwidth}
    \centering
    \includegraphics[width=\textwidth]{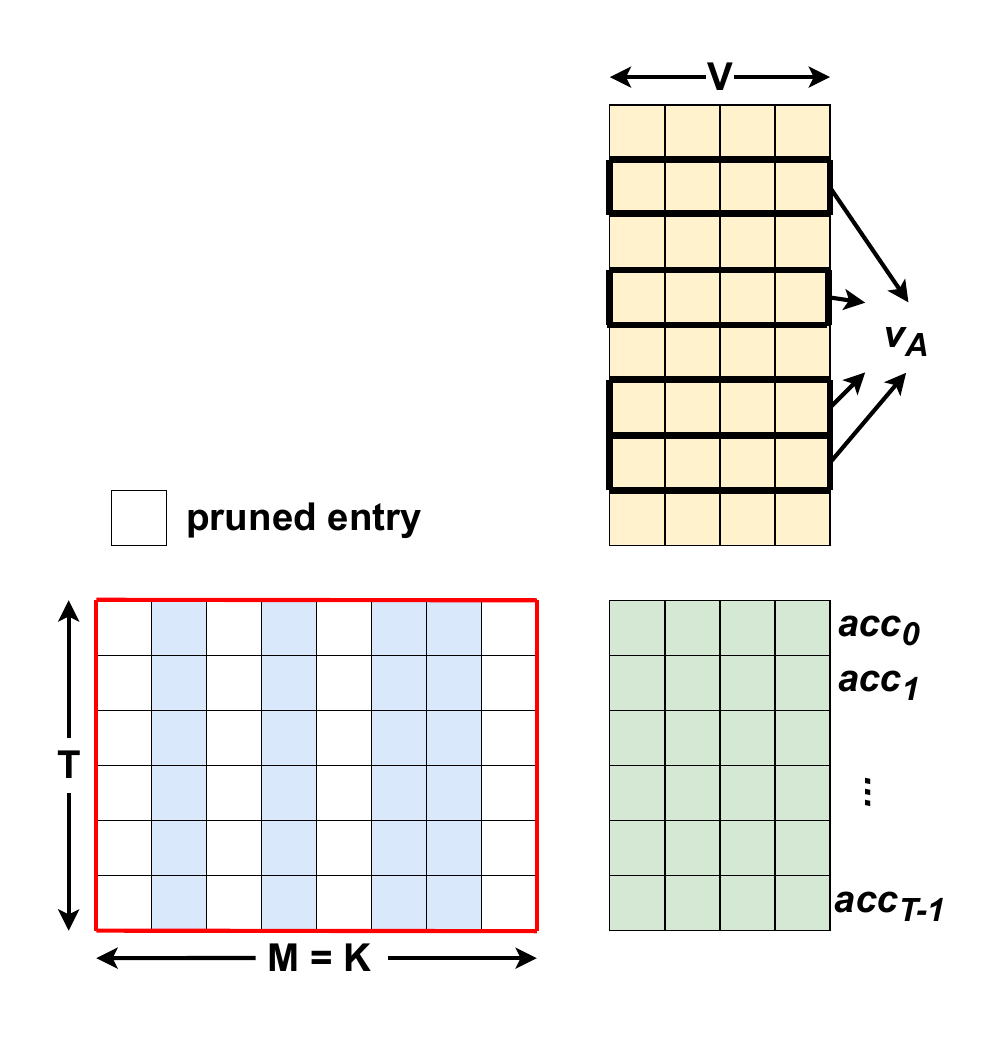}
    \caption{Our column-wise N:M pruning.\\(N=K/2 and M=K for 50\% sparsity)}
    \label{fig:ColumnWiseNMPruning}
    \Description{ColumnWiseNMPruning}
  \end{subfigure}
  \caption{Traditional N:M pruning and our proposed column-wise N:M pruning.}
  \label{fig:pruning}
\end{figure*}

The objective of this work is to accelerate CNN inference by leveraging N:M pruning.
Since convolutional layers dominate the computational cost of CNNs, we focus on optimizing their execution efficiency.
To this end, we employ the GEMM-based convolution technique that converts each convolution into a matrix multiplication.
The problem therefore reduces to optimizing the multiplication of a sparse weight matrix (pruned with N:M sparsity) and a data matrix, as illustrated in Figure \ref{fig:pruning}.
We further decompose this operation into tiled matrix multiplications, each executed by a high-performance RISC-V micro-kernel (Figure~\ref{fig:GEMM}).
The full matrix multiplication is then performed by iteratively invoking the micro-kernel across all output tiles.

Figure~\ref{fig:NMPruning} shows an example of a weight matrix pruned by the conventional N:M pruning method with N=2 and M=4.
Note that, in practice, the sparse weight matrix is stored in a compressed format to optimize memory usage.
However, the figure presents the original sparse layout for clarity.

To perform matrix multiplication under this pattern, we can utilize the inner-product-based approach.
This method iterates over each non-zero element in each {\em row} of the weight matrix.
Using the index matrix from the N:M compressed format, corresponding positions of the required columns in the dense data matrix are computed (e.g., the dashed box in Figure~\ref{fig:NMPruning}).
The retrieved elements are multiplied by the non-zero weights to form partial inner-products, which are then accumulated to produce one element of the output.
Since the data matrix is dense, we can further optimize the computation by simultaneously computing multiple elements from adjacent columns in a vectorized fashion, thus improving overall throughput.
This process is repeated for each row of the weight matrix.
However, due to indirect access to the data matrix, the required elements in the data matrix need to be reloaded, even if they were recently fetched for previous weight matrix rows.
This results in {\em redundant memory accesses} and potentially degrades performance, especially when cache locality is limited.

Alternatively, we can adopt an outer-product-based approach for matrix multiplication.
In this method, we iterate over the non-zero elements in each {\em column} of the weight matrix.
Since these weights all interact with the same element (or vector) of the data matrix, the data can be reused across multiple products, significantly reducing the redundant memory loads typically incurred by the inner-product method.
However, the non-zero weights within each column appear at irregular positions, causing the resulting partial products to be scattered across the output matrix. These partial results must be written back to memory and later reloaded for accumulation, introducing {\em redundant memory accesses}.

From the aforementioned outer-product-based approach, we observe that elements of the data matrix can be effectively reused if we process the non-zero weights along the column dimension of the weight matrix.
However, the irregular access patterns inherent to the N:M pruning structure still lead to redundant memory accesses when partial results are accumulated.
To address this, we propose  \textbf{column-wise N:M pruning}, which extends N:M pruning to a {\em column-wise} pruning strategy.
As illustrated in Figure \ref{fig:ColumnWiseNMPruning}, all the weights in a column are grouped as a unit, and each group is either entirely pruned or retained.
This strategy enforces a regular sparsity pattern, yielding two key advantages:
(1) it enhances reuses of data matrix rows across multiple weights, and (2) it enables accumulators to remain in registers rather than memory, thereby reducing both memory traffic and overall latency.

We use the L1 norm to evaluate the importance of each column group when deciding which to prune.
While column-wise pruning imposes stricter constraints compared to conventional row-wise N:M pruning, potentially affecting model accuracy, we mitigate this by increasing the group size M.
Since our approach targets CPU architectures, we can afford larger M values without adding computational overhead. 
In practice, M can span the entire input-channel dimension (i.e., a full row of the weight matrix).
For a given sparsity ratio, the number of retained columns is computed as  $N = \text{(1 - sparsity\_ratio)} \times M$.
This choice allows us to approximate unstructured pruning and maintain accuracy, while still preserving the structured sparsity necessary for efficient execution on general-purpose hardware.

Algorithm~\ref{alg:ColumnWiseNMPruning} presents the pseudo-code for the micro-kernel used in our tiled, column-wise matrix multiplication.
We begin by reserving $T$ vector registers  as accumulators for outer-product partial sums, where $T$ is the tile size (Lines 3-5).
For each weight matrix column, we load each weight element into a scalar register and fetch the corresponding values from the data matrix into one vector register (Line 7).
Each scalar weight is then multiplied by the vector register, and the product is accumulated in its dedicated accumulator register (Lines 8–11).
We repeat this procedure for all N columns of the weight matrix and finally write the accumulated results back to the output matrix.

Several parameters affect the micro-kernel’s efficiency. 
The tile size $T$ determines the number of vector registers reserved as accumulators throughout the micro-kernel’s execution, and one additional register holds the data matrix row fetched from memory.
Thus, increasing the tile size $T$ can improve efficiency with better reuse of the data register, but it also increases register pressure due to the increased number of accumulator registers.
Moreover, as discussed in Section ~\ref{sec:rvv}, a higher LMUL increases the vector length——allowing each instruction to process more data and improve throughput——but further reduces the number of available vector registers. 
In Section~\ref{subsec:method_integrate_xnnpack_ait}, we will explore how to optimally configure tile size and LMUL  to balance these trade-offs and achieve high-performance RISC-V kernel execution.

\begin{algorithm}[t]
\caption{Tiled Matrix Multiplication for Column-Wise N:M Pruning}
\label{alg:ColumnWiseNMPruning}
\begin{algorithmic}[1]
  \State \textbf{Input:}
  \begin{itemize}[nosep]
    \setlength\itemsep{0pt}
    \item $T$                              \Comment{Tile size}
    \item $V$                              \Comment{Vector length}
    \item $N$                              \Comment{Number of retained elements in a group}
    \item {\tt A}\,$[K,\,V]$               \Comment{Input data matrix after data packing}
    \item {\tt W}\,$[T,\,N]$               \Comment{Compressed weight matrix (column-wise $N\!:\!M$)}
    \item {\tt Idx}\,$[N]$            \Comment{Indices to the nonzero column groups}
  \end{itemize}

\State \textbf{Output:}
  \begin{itemize}[nosep]
    \setlength\itemsep{0pt}
    \item {\tt C}\,$[T,\,V]$               \Comment{Output matrix}
  \end{itemize}
  \vspace{1ex}

  \LeftComment{Reserve and initialize $T$ vector accumulators, $acc_0, ..., acc_{T-1}$}
  \For{$t = 0,\dots,T-1$}
    \State $acc_t \gets \mathbf{0}$
  \EndFor
  \vspace{0.75ex}  
  \For{$n = 0,\dots,N-1$} \Comment{Iterate over the $N$ retained columns}
    \State $v_A \gets \mathrm{vector\_load}\bigl(A + Idx[n] \times V)$ \Comment{Load the corresponding data vector}
    \For{$t = 0,\dots,T-1$} 
      \State $s_t \gets W[t,Idx[n]]$ \Comment{Load the weight}
      \State $acc_t \gets acc_t + s_t \times v_A$ \Comment{Multiply–accumulate into each accumulator
  \stepcounter{footnote}\footnotemark}
    \EndFor
  \EndFor
  \vspace{0.75ex}
  \For{$t = 0,\dots,T-1$} \Comment{Store results back to $C$}
    \State $\mathrm{vector\_store}\bigl(C + Idx[t] \times V,\,acc_t\bigr)$
  \EndFor
\end{algorithmic}
\end{algorithm}
\footnotetext{We use RISC-V instruction \codebox{vfmacc.vf vd, rs1, vs2, vm}\, to perform the scalar-vector multiplication:
  \(\mathrm{vd}[i] = \mathrm{rs1} \times \mathrm{vs2}[i] + \mathrm{vd}[i]\),
  where {\tt rs1} is a scalar, {\tt vs2} is a vector, and \texttt{vd} is the destination vector register.}

\subsection{Fusion of Im2col and Data Packing}\label{subsec:method_merging_im2col_packing}

\begin{figure}[t!]
\center
  \includegraphics[width=12cm, height=4.5cm]{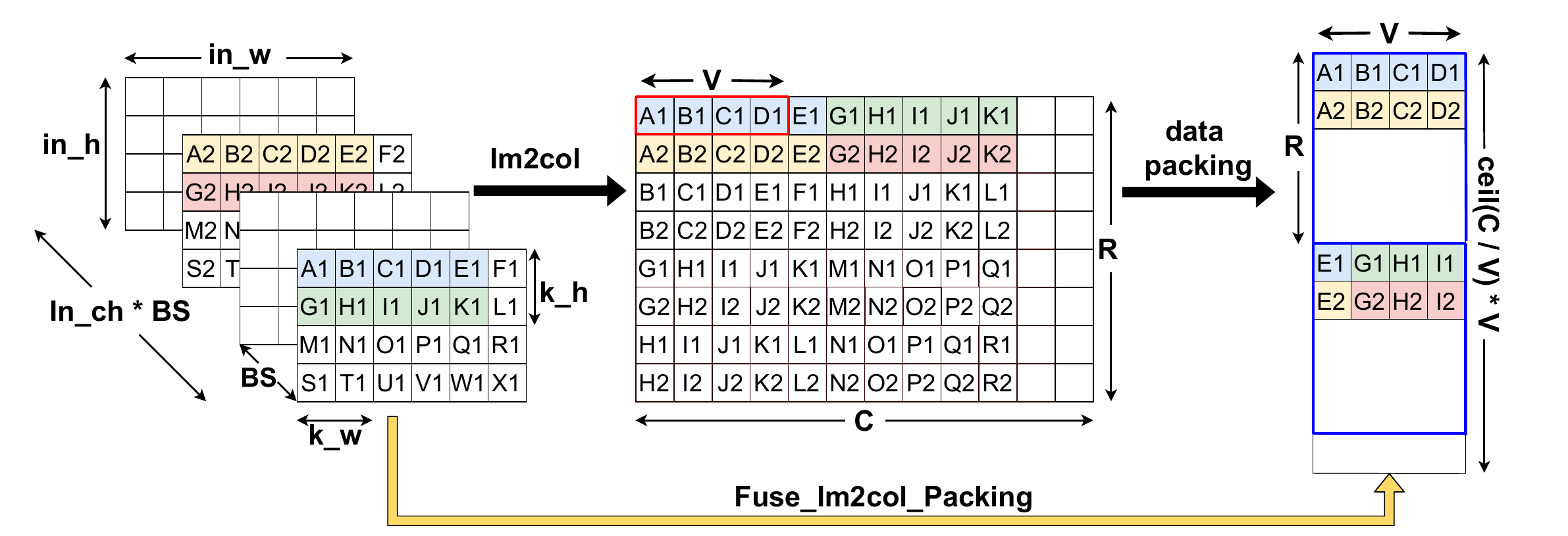}
  \vspace{-5mm}
  \caption{Illustration of fusing Im2col and data packing. Input feature map layout is CNHW, and kernel layout is OHWI.}
  \label{fig:Merging_Im2col_Packing}
  \Description{Diagram illustrating {\em Fuse\_Im2col\_Packing}.}
\end{figure}

Before performing matrix multiplication on our column-wise N:M pruned CNNs, two preprocessing steps are required.
First, we conduct {\em im2col} on the four-dimensional input feature maps, converting them into a two-dimensional matrix to leverage our efficient micro-kernel.
Next, we conduct {\em data packing} on the matrix, reorganizing it into a memory layout optimized for RISC-V vector execution and improved cache locality.
However, executing these two memory-intensive operations as separate steps introduces significant memory overhead. To fully realize the performance benefits of our approach, minimizing this overhead is critical.

To address this, we integrate the im2col and data packing operations into a single memory-efficient step.
In our implementation, the CNN feature maps follow the CNHW layout, where C, N, H, and W  denote the input channels, batch size, input height, and input width, respectively.
As illustrated in Figure~\ref{fig:Merging_Im2col_Packing}, the im2col transformation performs a sliding window convolution over the H and W dimensions (scanning the W dimension first), and reorganizes the feature maps into columns of the resulting data matrix.
Since W is the innermost dimension in the CNHW layout, elements in the W dimension are stored contiguously in memory.
This property enables efficient vectorization: we can use vector instructions to load multiple contiguous elements simultaneously and write them directly to the matrix.
Concurrently, we apply data packing to align the matrix layout with the RISC-V vector architecture.
Specifically, we reorganize the data into vector-aligned {\em strips}, each with a size equal to the vector length (as shown in Figure~\ref{fig:Data_Packing}).
This design enables us to move data directly from the feature maps into strips in a single pass, eliminating the need for two separate memory operations.
For each strip, we compute the source data address in the feature maps, and issue a single vector instruction to transfer the relevant elements efficiently. 

A challenge in vector-based data movement arises when the input feature map width is not a multiple of the vector length, which is typically a power of two. registers are typically sized as powers of two.
For example, if the input width is 56 and the vector length is 32, there will be 24 elements remaining to be included in a strip.
In fixed-length SIMD architectures (e.g., x86 AVX), this issue is usually handled via masked load/store instructions to avoid invalid memory access.
In contrast, scalable vector architectures like RISC-V, which support a flexible vector length (VL), allow us to dynamically adjust VL to match the number of remaining elements.  This flexibility not only ensures safe memory access but also avoids unnecessary data movement, such as copying zero-padding regions, thereby further enhancing efficiency.

Furthermore, using a larger vector length (i.e., larger LMUL) increases the number of elements processed per instruction, reducing loop iteration overhead and improving data movement efficiency.
However, if the number of data elements to be processed is small (e.g., short input width), it may lead to under-utilization of vector registers and  degrade performance.
To address this, in our fused im2col and data packing micro-kernel, we dynamically adjust the LMUL value to make the vector length more closely aligned with the feature map width.
This adaptive strategy maximizes vector register utilization and ensures high efficiency regardless of input shape. 
Our evaluation confirms that selecting an LMUL that yields a vector length close to the input width consistently delivers optimal performance.
The pseudo-code for our fused im2col and data packing operation is presented in Algorithm~\ref{alg:merging_im2col_packing}.

\vspace{-2.5mm}
\begin{algorithm}[H]
\caption{Fused Im2col and Data Packing}
\label{alg:merging_im2col_packing}
\begin{algorithmic}[1]
\State \textbf{Input:}
\begin{itemize}[nosep]
      \item $B$                \Comment{batch size}
      \item $C_{\mathrm{in}}$  \Comment{number of input channels}
      \item $H_{\mathrm{in}},W_{\mathrm{in}}, H_{\mathrm{out}},W_{\mathrm{out}}$ \Comment{input/output  height/width}
      \item $K_h,K_w$          \Comment{kernel height/width}
      \item $padding$   \Comment{padding}
      \item $V$               \Comment{Vector length}
      \item {\tt input}\;$[C_{\mathrm{in}},B,H_{\mathrm{in}},W_{\mathrm{in}}]$
      \item {\tt Im2col\_Packing\_Out}\,$[\lceil\frac{B\,H_{\mathrm{out}}\,W_{\mathrm{out}}}{V}\rceil,\;K_hK_wC_{\mathrm{in}},\;V]$
    \end{itemize}
\State $batch\_output\_size \gets B \times H_{\mathrm{out}} \times W_{\mathrm{out}}$
\For{$out_{\mathrm{cur}} = 0,\dots,batch\_{\mathrm{output\_size}}-1$ \textbf{step} $V$}
    \For{$k = 0$ \textbf{to} $K_hK_w-1$}   \Comment{traverse all kernel elements}
        \For{$in\_ch = 0$ \textbf{to} $C_{\mathrm{in}}$}
            \For{$cur = 0$ \textbf{to} $V$ \textbf{step} $\min(V, W_{\mathrm{out}})$}
                \State $vl \gets \min(V, W_{\mathrm{out}}) - padding$
                \State // {`input\_cur' points to the start of the input data to load}
                \State $va \gets \mathrm{vector\_load}(input\_cur, vl)$
                \State // {Store the loaded vector into output matrix at specified offset}
                \State $\mathrm{vector\_store}(\textit{Im2col\_Packing\_Out} + output\_offset, va, vl)$ 
            \EndFor
        \EndFor
    \EndFor
\EndFor
\end{algorithmic}
\end{algorithm}
\vspace{-5.5mm}

\subsection{The Auto-tuning Framework}\label{subsec:method_integrate_xnnpack_ait}

We design our column-wise N:M pruning optimization in a framework that integrates XNNPACK and the AITemplate AI compiler.
AITemplate originally supports the backends of NVIDIA and AMD GPUs only.
Therefore, we extend AITemplate with the XNNPACK backend to support execution on RISC-V CPUs.

AITemplate is an AI compiler that transforms machine learning models into optimized executable.
For each operator in the model, it first generates a set of kernel candidates in C++.
The kernel code interface is parameterized by utilizing C++ templates, allowing developers to customize the kernels by specifying various template parameters, such as tile sizes, unrolling factors, and memory layouts.
AITemplate then profiles the execution performance of these kernel candidates on the target hardware, and selects the fastest one as the operator's implementation to be included in the final optimized executable.

In this work, we implement our optimization strategies as XNNPACK micro-kernels.
As previously discussed, two key parameters significantly impact micro-kernel efficiency: (1) the tile size $T$, which determines the number of vector registers reserved as accumulators, and (2) LMUL, which controls the effective vector length.
Increasing either parameter can improve register reuse and parallelism, but excessive values may lead to register pressure and diminished returns.
To explore optimal configurations, we parameterize our micro-kernels using templates, allowing us to leverage AITemplate’s profiling mechanism for automatic selection of the most efficient kernel implementation. 

For the tile size T, we profile values from 1 to 32, matching the total number of vector registers available in RISC-V. 
For LMUL, although RISC-V supports values of 1/8, 1/4, 1/2, 1, 2, 4, and 8, our experiments reveal that smaller LMUL values (1/8, 1/4, and 1/2) reduce vector parallelism and degrade performance. Therefore, we restrict our profiling to LMUL values of 1, 2, 4, and 8 to ensure optimal efficiency.

\section{Performance Evaluation}\label{sec:experiment}

In this section, we present a comprehensive evaluation of our optimizations.
We begin by describing the experimental settings.
Then, we evaluate the execution efficiency of our column-wise N:M pruning and fusion optimization. Finally, we report the end-to-end results and model accuracy on several CNN architectures.

\subsection{Experimental Settings}\label{subsec:expr_setup}
\subsubsection{Platform and Profiling Setup}\

All benchmarks are executed on a Banana Pi BPI-F3 board ~\cite{banana-pi-f3} equipped with a SpacemiT K1 8-core RISC-V CPU and 8 GB LPDDR4x memory.
The CPU features the RVV 1.0 extension with a base vector length of 256 bits.
The system runs Bianbu OS with Linux kernel v6.1.15.
XNNPACK and AITemplate are compiled using Clang 17.0.2 with optimization level -O3 and RVV support.
We use the SiFive‑optimized RVV branch of XNNPACK~\cite{sifive_xnnpack_rvv} as the dense NHWC baseline.
In the experiments of Section 4.2 and 4.3, we evaluate single-thread performance, and in the rest of Sections, we use multithreading to process output tiles in parallel, using the default XNNPACK setting.
Performance metrics of L1-cache loads are collected using the Linux performance monitoring tool \texttt{perf}.
Unless explicitly stated, all experiments use a batch size of 1.

\subsubsection{Models and Training}
Models used in this study include ResNet-18, ResNet-34, ResNet-50, ResNet-101, ResNet-152~\cite{he2016deep}, MobileNet-V2~\cite{sandler2018mobilenetv2}, and DenseNet-121~\cite{huang2017densely}.
All models are downloaded from PyTorch's Torchvision model zoo.
We prune the models using one-shot pruning and retrain them on the ImageNet dataset~\cite{deng2009imagenet} for 90 epochs to recover the accuracy using an NVIDIA RTX 3090 GPU.
Model retraining employs the AdamW optimizer.
Specifically, DenseNet-121 follows the standard training protocol by PyTorch.
ResNet models are trained with an initial learning rate of $10^{-4}$, which decays tenfold every 30 epochs.
MobileNet-V2 adopts the same initial learning rate but decays at epochs 30, 65, and 85.

We do not prune the first convolution layer, as it has only 3 input channels and contributes marginally to the overall computational cost.
Since our approach operates on the CNHW data layout and the models' default layout is NHWC, we transform the layout from NHWC to CNHW before the first convolution.
The CNHW layout is then used throughout the model, and the output is converted back to the original NHWC layout after the last convolution layer.

\subsection{Inference Time Evaluation of Convolution Layers}

\begin{figure*}[t!]
\centering
\begin{tikzpicture}[scale=0.6]
\begin{axis}[
    width=20cm,
    height=6cm,
    ymin=0,
    ymax=120,
    enlarge y limits=false,
    ybar=0cm,
    bar width=0.4cm,
    enlarge x limits={abs=1cm},
    xtick=data,
    tick label style={font=\Large},
    symbolic x coords={
          Stage1-conv1, Stage1-conv2, Stage1-conv3, 
          Stage2-conv1, Stage2-conv2, Stage2-conv3,
          Stage3-conv1, Stage3-conv2, Stage3-conv3,
          Stage4-conv1, Stage4-conv2, Stage4-conv3,
      },
      ylabel={\parbox{5cm}{\centering \huge Runtime (ms)}},
      legend style={at={(0.46,1.2)}, anchor=north, font=\LARGE},
      x tick label style={rotate=30},
      legend columns=3, 
      legend cell align=left, 
    ]
      \addplot [draw=black, fill=white, ultra thick, area legend] 
        coordinates {
          (Stage1-conv1,8.87705)
          (Stage1-conv2,35.3667)
          (Stage1-conv3,17.3051)
          (Stage2-conv1,7.68495)
          (Stage2-conv2,25.4186)
          (Stage2-conv3,11.4396)
          (Stage3-conv1,9.35998)
          (Stage3-conv2,22.3739)
          (Stage3-conv3,15.1014)
          (Stage4-conv1,15.1056)
          (Stage4-conv2,21.6714)
          (Stage4-conv3,22.0348)
        };
      \addplot [draw=black, fill=gray, ultra thick, area legend] 
        coordinates {
          (Stage1-conv1,12.4543)
          (Stage1-conv2,82.2158)
          (Stage1-conv3,32.3818)
          (Stage2-conv1,12.3929)
          (Stage2-conv2,74.2303)
          (Stage2-conv3,27.5195)
          (Stage3-conv1,13.7411)
          (Stage3-conv2,77.2784)
          (Stage3-conv3,24.8582)
          (Stage4-conv1,17.4016)
          (Stage4-conv2,118.14)
          (Stage4-conv3,38.3325)
        };
      \addplot [draw=black, fill=red!50, pattern=north east lines, ultra thick, area legend, nodes near coords={
             \pgfmathprintnumber[fixed,precision=1]{\pgfplotspointmeta}
          },
          every node near coord/.append style={font=\large, rotate=45, anchor=west}] 
        coordinates {
          (Stage1-conv1,5.07169)
          (Stage1-conv2,28.013)
          (Stage1-conv3,10.2494)
          (Stage2-conv1,4.11179)
          (Stage2-conv2,18.8943)
          (Stage2-conv3,6.37204)
          (Stage3-conv1,8.07191)
          (Stage3-conv2,19.2854)
          (Stage3-conv3,8.31537)
          (Stage4-conv1,8.90589)
          (Stage4-conv2,16.087)
          (Stage4-conv3,13.3795)
        };
\legend{Dense, Conventional N:M Pruning, Column-Wise N:M Pruning}
\end{axis}
\end{tikzpicture}
\vspace{-6mm}
\caption{Comparison of inference time for the convolution layers in ResNet-50.}

\label{fig:Pruning_Runtime}
\Description{Runtime Comparison of GEMM microkernel Implementations.}
\vspace{-2mm}
\end{figure*}
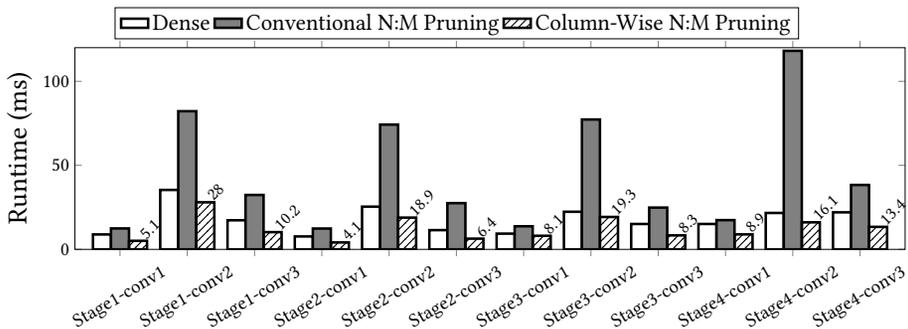

In this experiment, we evaluate the inference time of convolution layers in ResNet-50 as a case study.
ResNet-50 comprises four stages, each containing three representative convolution layers.
We select these layers with varying shapes for evaluation, excluding the downsampling layers.
We compare three configurations: (1) dense, (2) conventional N:M pruning using an outer-product-based scheme, and (3) our column-wise N:M pruning method.
All three configurations apply our fused im2col and data packing optimization and use the CNHW layout.
For pruned models, we evaluate at 50\% sparsity.
As Figure~\ref{fig:Pruning_Runtime} shows, the conventional N:M pruning method exhibits significant slowdown, with up to \(5.4\times\) performance degradation compared to the dense baseline. 
This decline is primarily due to memory overhead from redundant memory accesses. 
In contrast, our column-wise method consistently outperforms the dense baseline, achieving up to a \(1.86\times\) speedup and an average of 1.5$\times$.
These results demonstrate the effectiveness of our column-wise pruning design over the outer-product-based approach. 

\subsection{Performance of Fusing Im2col and Data Packing}\label{subsec:expr_merging_im2col_packing}

\begin{figure}[t!]
\center
  \includegraphics[width=12cm, height=3.5cm]{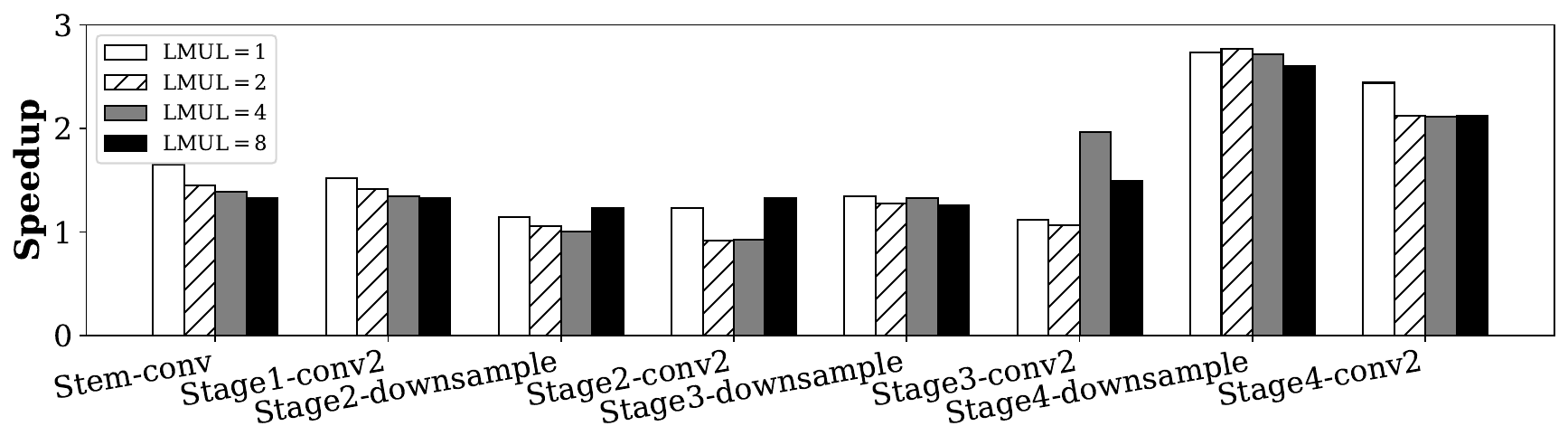}
  \vspace{-7mm}
  \caption{Speedup of our fusion optimization compared to performing im2col and data packing separately, under different LMUL configurations for ResNet-50.}
  \label{fig:Speedup_merging_im2col_packing}
  \vspace{-3mm}
\end{figure}

\begin{figure}[t!]
\center
  \includegraphics[width=9cm, height=3.5cm]{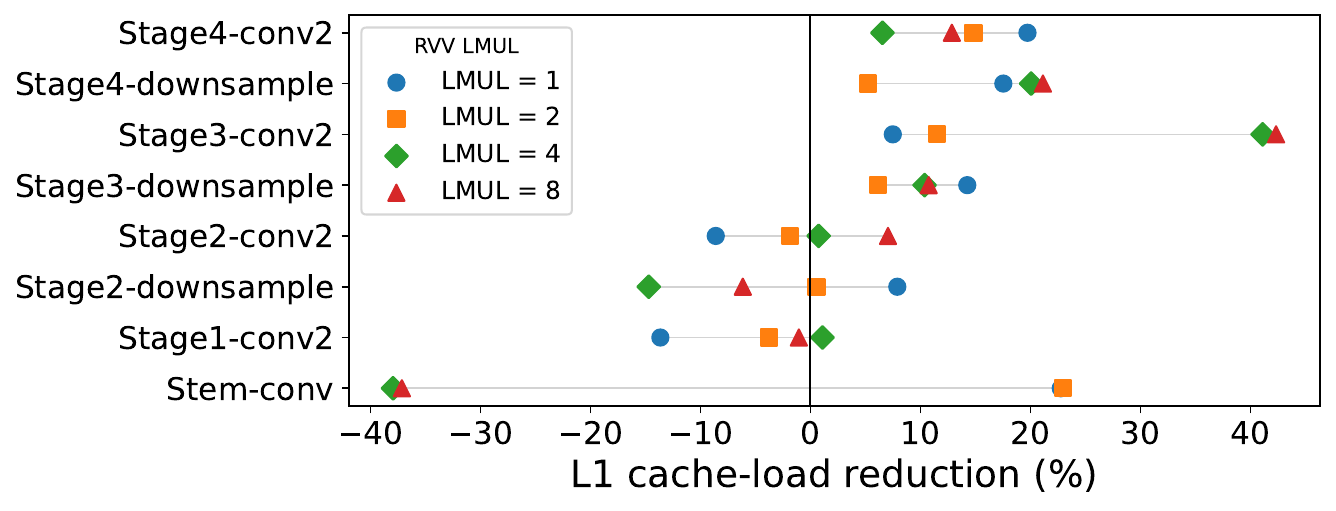}
  \vspace{-3mm}
  \caption{Reduction of L1‑cache loads by fusing im2col and data packing over the non-fused approach, across various LMUL configurations for 3x3 convolution layers in ResNet-50.}
  \label{fig:L1_cache_load_compare}
  \Description{L1 Cache Load Reduction}
\end{figure}

This section evaluates the performance of input feature map preprocessing using our fused im2col and data packing optimization, compared to a baseline that performs im2col and data packing separately.
We examine the stem layer and the second convolution layers of each stage in ResNet-50, as these layers use 7x7 and 3x3 convolutional kernels that contribute significantly to im2col overhead.
Figure~\ref{fig:Speedup_merging_im2col_packing} show the performance speedup across different LMUL values.
The result indicates that fusion consistently outperforms the baseline regardless of the LMUL settings, due to reduced memory access overhead.
We further observe that the optimal LMUL value varies across layers. This variation stems from the input feature map widths not being divisible by the vector length, requiring multiple instructions to handle edge elements. While increasing LMUL improves memory transfer efficiency via longer vectors, it also raises overhead for handling boundary cases, which may limit overall performance gains.
Figure~\ref{fig:L1_cache_load_compare} shows the reduction in L1 cache loads for various LMUL values compared to the baseline. The result suggests a strong correlation between greater speedup (Figure~\ref{fig:Speedup_merging_im2col_packing}) and a larger reduction in L1 cache accesses.


\begin{figure}[t]
  \centering
  \begin{subfigure}[b]{0.48\textwidth}
    \centering
    \includegraphics[width=\linewidth, height=3.6cm]{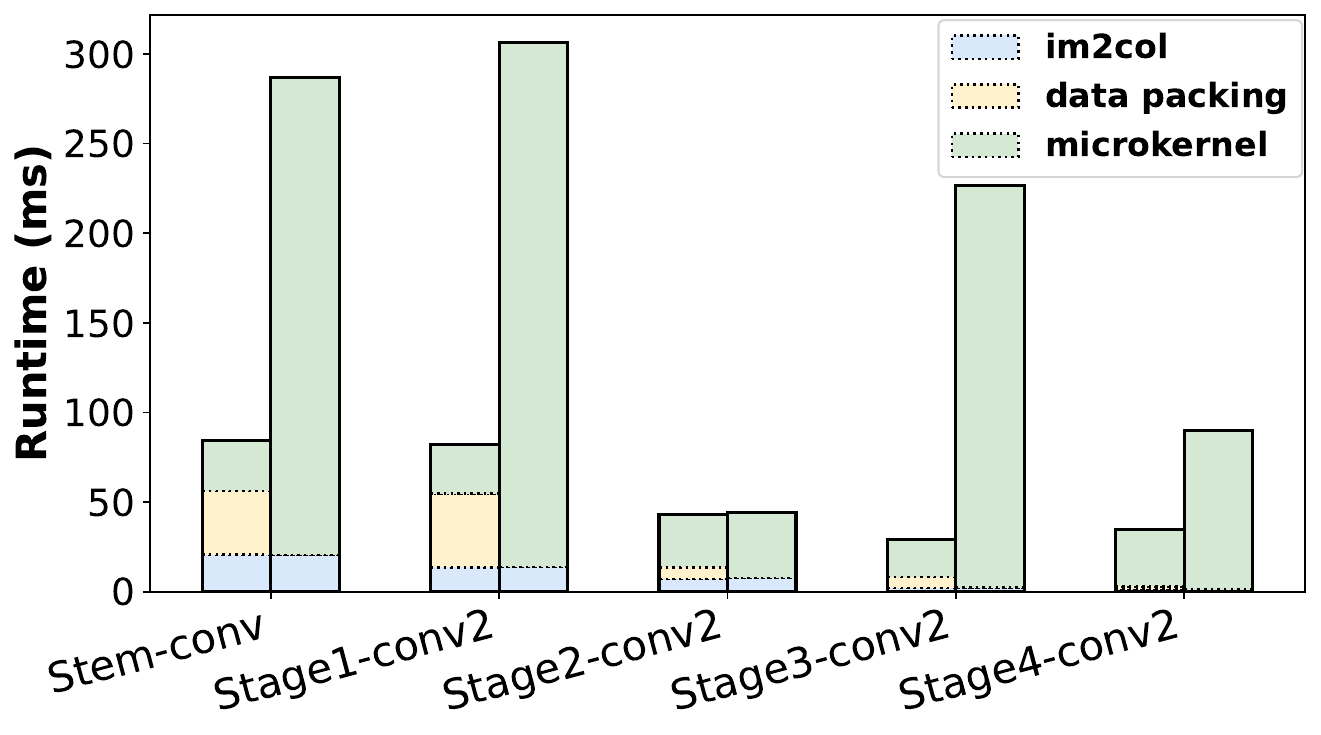}
    \caption{Comparison with and without data packing.}
    \label{fig:runtime_dense_nopack}
  \end{subfigure}\hfill
  \begin{subfigure}[b]{0.48\textwidth}
    \centering
    \includegraphics[width=\linewidth, height=3.6cm]{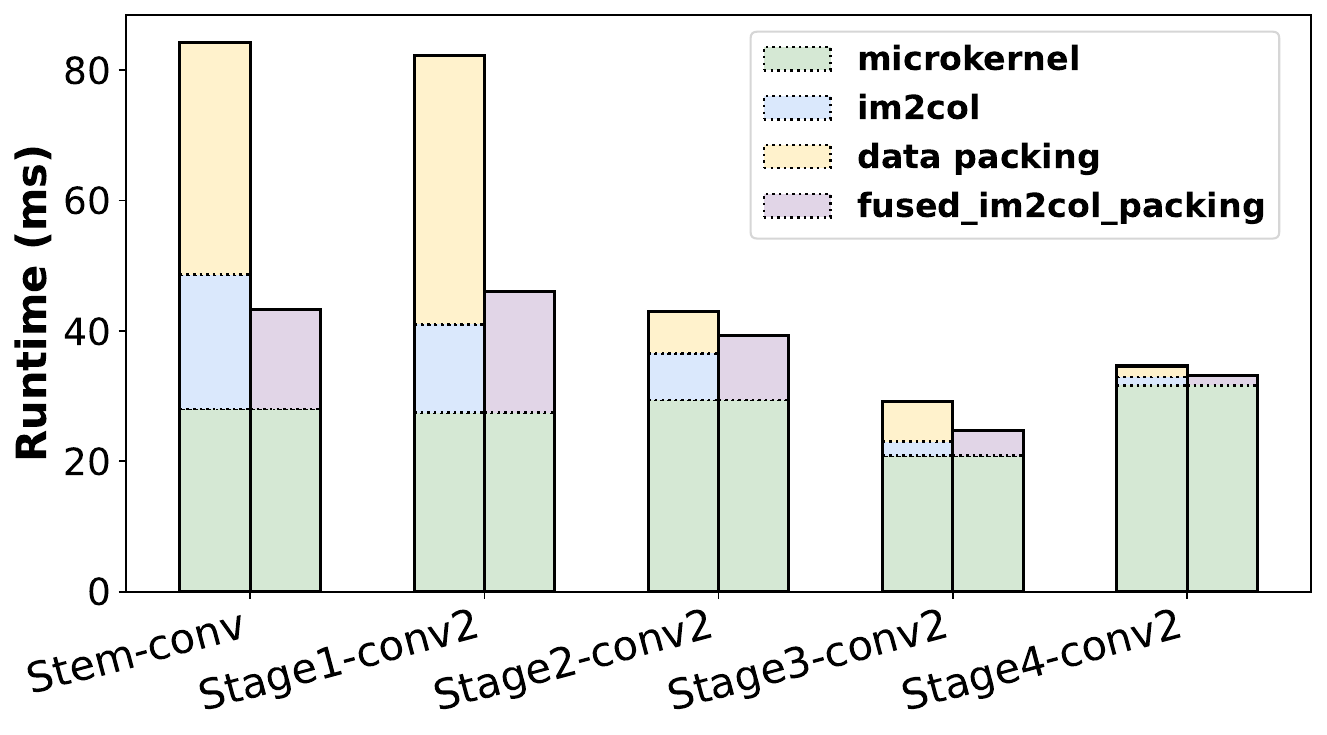}
    \caption{Comparison with and without our fusion optimization.}
    \label{fig:runtime_dense_fused}
  \end{subfigure}

  \caption{Execution time breakdown.}
  \vspace{-3mm}
  \label{fig:runtime_breakdown_two_plots}
\end{figure}

To further highlight the necessity of our proposed optimization, we break down the convolution inference time for three configurations: (1) performing im2col and data packing as separate operations, (2) performing im2col only (no data packing), and (3) our fused im2col and data packing approach.
We use unpruned models in this evaluation to isolate the overhead from sparsity-related optimizations.

Figure~\ref{fig:runtime_dense_nopack} compares the execution time with and without data packing.
We observe that disabling data packing significantly increases overall execution time, due to the significant rise in the execution time of the matrix multiplication microkernel.
This performance degradation results from poor cache locality when data packing is omitted, highlighting data packing as essential for efficient matrix multiplication. 
Figure~\ref{fig:runtime_dense_fused} shows the execution time with and without our fusion optimization.
The result shows that fusing im2col with data packing only slightly increases the execution time against the im2col operation alone, while significantly reducing the time over performing the two operations separately.
Notably, for the Stem-conv layer, the fused method even surpasses im2col alone in speed. 
This convolution layer uses a stride of 2, and when im2col and data packing are decoupled, extra padding is needed, introducing additional memory overhead. 
In contrast, our method intelligently adjusts memory offsets to avoid these padded regions, thus minimizing the memory overhead.
Overall, these results show the importance of data packing for improving cache locality and demonstrate that our fusion optimization is essential for achieving full performance benefits.

\subsection{Multi-thread Performance of Convolution Layers}\label{subsec:expr_conv2d_lmul}

\begin{figure}[t!]
\center
  \includegraphics[width=12cm, height=5cm]{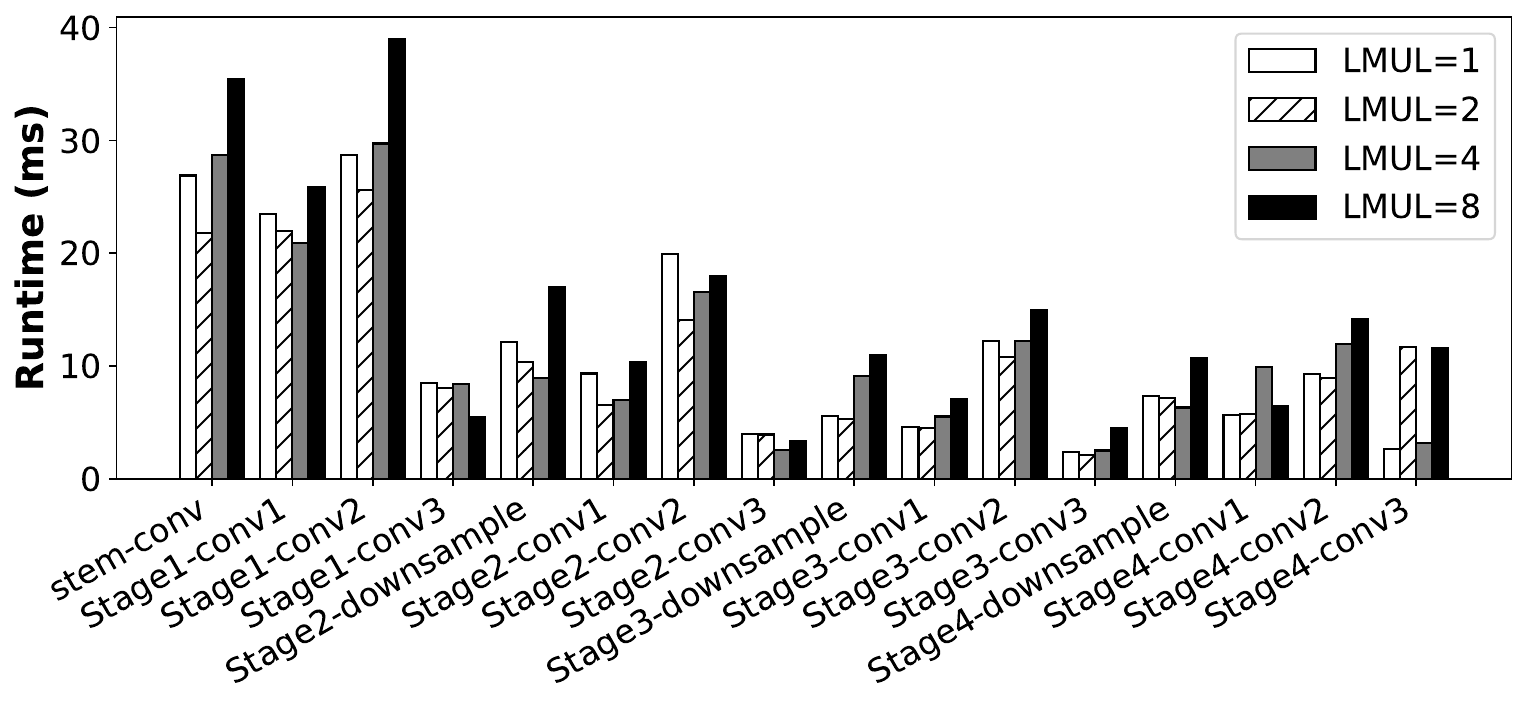}
  \vspace{-7.5mm}
  \caption{Convolution inference time across LMUL values with our column-wise N:M pruning.}
  \label{fig:Conv2d_with_pruning_different_lmul}
\end{figure}

This section evaluates the execution efficiency of convolution layers in ResNet-50 under multi-threaded execution with 50\% sparsity.
Figure~\ref{fig:Conv2d_with_pruning_different_lmul} shows the inference time results for different LMUL configurations: 1, 2, 4, and 8.
The results show that performance varies across layers depending on the LMUL value: e.g., LMUL=4 yields the best performance for Stage1-conv1, LMUL=2 for Stage1-conv2, and LMUL=8 for Stage1-conv3.
The optimal LMUL can yield up to a 4$\times$ speedup compared to the worst-performing configuration.
This variability highlights that a static LMUL configuration is inadequate for optimal performance. Therefore, the tuning mechanism introduced in Section~\ref{subsec:method_integrate_xnnpack_ait} is critical for identifying the optimal LMUL configuration.

\begin{figure}[t!]
\center
  \includegraphics[width=12cm, height=6.5cm]{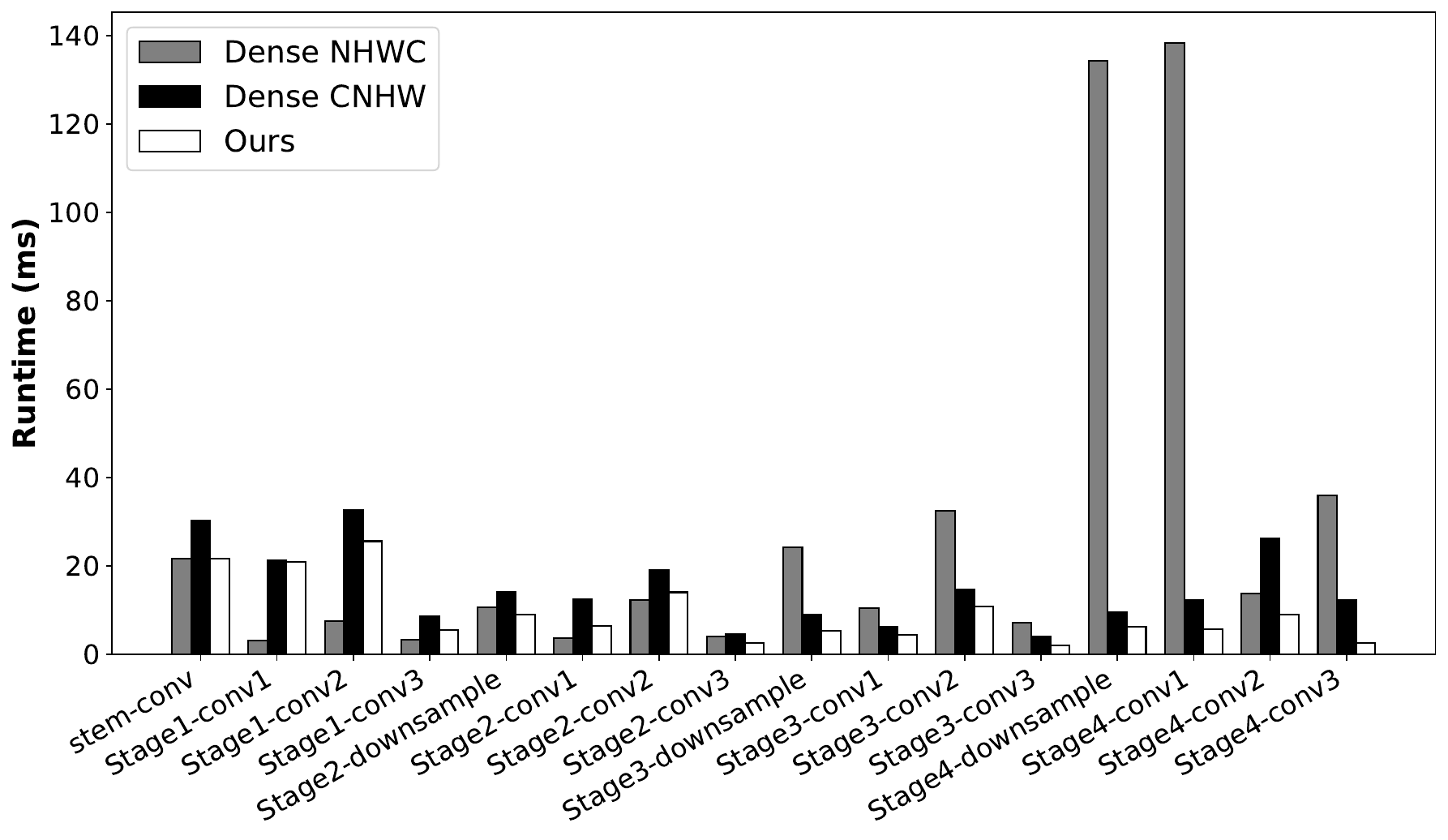}
  \caption{Inference time comparison with dense versions and our column-wise N:M pruning.}
  \label{fig:Conv2d_with_best_pruning_ver_vs_dense}
\end{figure}

Next, we compare the execution efficiency of our pruning method against two dense baselines.
The first baseline is a SiFive-optimized XNNPACK implementation, which uses the dense NHWC layout.
The second baseline uses the dense CNHW layout.
For both dense baselines, we fix LMUL to 4, matching the setting used in the SiFive implementation. 
In contrast, the pruned model uses our auto-tuning mechanism to select the optimal LMUL per layer.
As shown in Figure~\ref{fig:Conv2d_with_best_pruning_ver_vs_dense}, our pruning method consistently outperforms the dense CNHW baseline, achieving up to a 2.1$\times$ speedup.
In early layers (e.g., Stage 1), the dense NHWC baseline outperforms both the dense CNHW and our sparse CNHW configurations. 
However, its performance deteriorates significantly in deeper layers (e.g., Stages 3 and 4).
Notably, in layers such as Stage4-downsampling and Stage4-conv1, the SiFive implementation is up to 21$\times$ slower than our method.
This degradation stems from the memory handling strategy used in SiFive’s XNNPACK implementation, which performs data packing on the weight matrix and uses an indirection buffer for accessing input data.
As the weight tensors increase in size in deeper layers, this approach incurs substantial data movement overhead due to data packing, leading to significantly reduced performance.


\subsection{End-to-end Model Performance and Accuracy}

This section evaluates the end-to-end performance and ImageNet top-1 accuracy across varying sparsity levels and batch sizes.
We compare the following configurations:
\begin{enumerate}[itemsep=0pt,nosep]
\item Conventional row-based N:M pruning with fixed N and M=4; equivalent to our column-wise method with a tile size of 1.
\item Column-wise N:M pruning with fixed N and M=4, using a tile size of 8. 
\item Column-wise N:M pruning with adaptive N and M, using a tile size of 8. The value of M is adjusted based on the number of input channels  per convolution layer. 
\item Column-wise N:M pruning with adaptive N, M, and tile size. The value of M is adjusted based on the number of input channels for each convolution layer, and the tile size is determined using our auto-tuning mechanism.
\end{enumerate}

\begin{table}[t]
\centering
\caption{Comparison of ImageNet top-1 accuracy across various pruning patterns for ResNet-50. \textbf{T} denotes the tile size.}
\label{tab:Resnet50_accuracy}
\vspace{-2mm}
\begin{tabular}{|c|c|c|}
\hline
\textbf{Sparsity} & \textbf{Variant} & \textbf{Top-1 Accuracy}\\
\hline
\multirow{1}{*}{Dense} 
& Dense  & 76.1\% \\
\hline
\multirow{4}{*}{
    \parbox[c]{3cm}{\centering
        25\%~Sparsity%
    }
} 
& 3:4 (T = 1)                                & 75.9\% \\
& 3:4 (T = 8)                                & 75.8\% \\
& columnwise N:M pruning (T = 8)  & 75.8\% \\
& columnwise N:M pruning           & 75.7\% \\
\hline
\multirow{4}{*}{
    \parbox[c]{3cm}{\centering
        50\%~Sparsity%
    }
}
& 2:4 (T = 1)                                & 76.0\% \\
& 2:4 (T = 8)                                & 74.3\% \\
& columnwise N:M pruning (T = 8)  & 75.4\% \\
& columnwise N:M pruning           & 75.4\% \\
\hline
\multirow{4}{*}{
    \parbox[c]{3cm}{\centering
        75\%~Sparsity%
    }
}
& 1:4 (T = 1)                                & 74.7\% \\
& 1:4 (T = 8)                                & 73.7\% \\
& columnwise N:M pruning (T = 8)  & 73.9\% \\
& columnwise N:M pruning           & 74.2\% \\
\hline
\end{tabular}
\end{table}
\begin{figure}[t!]
\center
  \includegraphics[width=12cm, height=5cm]{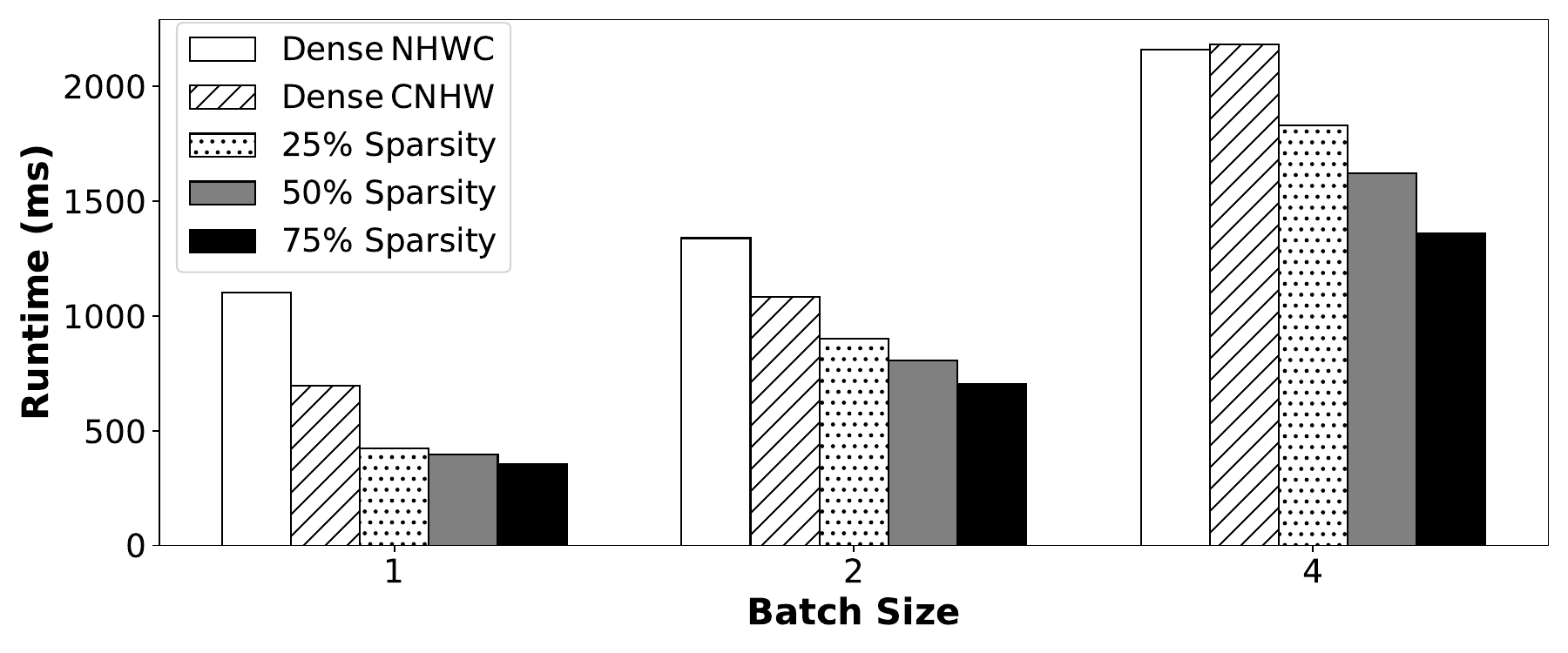}
  \vspace{-9mm}
  \caption{Comparison of ResNet‑50 inference time across various batch sizes and sparsity levels.}
  \label{fig:resNet50_diff_bs}
  \Description{ResNet50 with different bs, sparsity}
  \vspace{-4mm}
\end{figure}

Table~\ref{tab:Resnet50_accuracy} reports the top-1 accuracy of ResNet-50 under different configurations. 
The conventional row-based method (Configuration 1) yields the highest accuracy, as it imposes fewer structural constraints than column-wise pruning. 
This difference is evident when comparing Configuration 1 and 2, where both use the same N and M values, but the added column-wise constraint in Configuration 2 leads to a noticeable accuracy drop.
However, when N and M are adaptively chosen based on each layer’s channel count (Configurations 3 and 4), the resulting sparsity patterns resemble unstructured pruning, leading to a significant accuracy recovery.
Across all tested settings, accuracy degradation remains below 1\% for sparsity levels of 25–50\% and below 2\% at 75\% sparsity, achieving performance comparable to conventional row-wise N:M pruning.

\begin{table}[tbp]
  \centering
  \caption{Top-1 accuracy (Acc) and inference time (Time, ms) for dense models and sparse models (r = sparsity\_ratio) on the ImageNet dataset.}
  
  \label{tab:model_accuracy_runtime_7subtables}
  \begin{subtable}[t]{0.5\textwidth}
    \centering
    \begin{tabular}{@{}lrr@{}}
      \toprule
      ResNet-18 & Time & Acc \\
      \midrule
      Dense   & 649.3 & 69.7\% \\
      r=0.25 & 200.8 & 70.0\% \\
      r=0.50 & 180.5 & 69.5\% \\
      r=0.75 & 165.1 & 67.6\% \\
      \bottomrule
    \end{tabular}
  \end{subtable}\hfill
  \begin{subtable}[t]{0.5\textwidth}
    \centering
    \begin{tabular}{@{}lrr@{}}
      \toprule
      ResNet-34 & Time & Acc \\
      \midrule
      Dense   & 1205.1 & 73.3\% \\
      r=0.25 & 401.0 & 73.4\% \\
      r=0.50 & 348.9 & 73.0\% \\
      r=0.75 & 300.0 & 71.9\% \\
      \bottomrule
    \end{tabular}
  \end{subtable}


  \mycomment{
  \begin{subtable}[t]{0.5\textwidth}
    \centering
    \begin{tabular}{@{}lrr@{}}
      \toprule
      ResNet-50 & Time & Acc \\
      \midrule
      Dense   & 1103.8 & 76.1\% \\
      r=0.25 & 424.9 & 75.7\% \\
      r=0.50 & 395.0 & 75.4\% \\
      r=0.75 & 356.6 & 74.2\% \\
      \bottomrule
    \end{tabular}
  \end{subtable}
  }
  \hfill
  
  \begin{subtable}[t]{0.5\textwidth}
    \centering
    \begin{tabular}{@{}lrr@{}}
      \toprule
      ResNet-101 & Time & Acc \\
      \midrule
        Dense   & 1968.7  & 77.3\% \\
        r=0.25 & 750.6   & 76.5\% \\
        r=0.50 & 705.0   & 76.3\% \\
        r=0.75 & 603.2   & 75.9\% \\
      \bottomrule
    \end{tabular}
  \end{subtable}\hfill
  \begin{subtable}[t]{0.5\textwidth}
    \centering
    \begin{tabular}{@{}lrr@{}}
      \toprule
      ResNet-152 & Time & Acc \\
      \midrule
        Dense & 2579.7   & 78.3\% \\
        r=0.25 & 1127.4   & 77.6\% \\
        r=0.50 & 1009.8   & 77.4\% \\
        r=0.75 & 898.6    & 77.0\% \\
      \bottomrule
    \end{tabular}
  \vspace{0.25em}

  \end{subtable}
  \begin{subtable}[t]{0.48\textwidth}
    \centering
    \begin{tabular}{@{}lrr@{}}
      \toprule
      MobileNet-v2 & Time & Acc \\
      \midrule
      Dense   & 100.6 & 71.8\% \\
      r=0.50 & 72.7 & 67.0\% \\
      r=0.75 & 68.0 & 59.8\% \\
      \bottomrule
    \end{tabular}
  \end{subtable}\hfill
  \begin{subtable}[t]{0.48\textwidth}
    \centering
    \begin{tabular}{@{}lrr@{}}
      \toprule
      DenseNet-121 & Time & Acc \\
      \midrule
        Dense   & 395.2   & 74.4\% \\
        r=0.50 & 377.7   & 75.0\% \\
        r=0.75 & 363.5   & 73.5\% \\
      \bottomrule
    \end{tabular}
  \end{subtable}
\end{table}

Figure~\ref{fig:resNet50_diff_bs} compares the ResNet-50 inference time across three configurations: (1) dense model using the NHWC layout (SiFive-optimized XNNPACK), (2) dense model using the CNHW layout, and (3) our sparse model at 25\%, 50\%, and 75\% sparsity.
At batch sizes 1 and 2, the dense CNHW layout outperforms NHWC due to reduced overhead. 
However, the performance gap narrows at batch size 4 due to increased memory movement. 
Across all batch sizes, our sparse models outperform dense counterparts. 
At 75\% sparsity, our pruning method achieves speedups of 3.0$\times$, 1.9$\times$, and 1.5$\times$ over the dense NHWC baseline for batch sizes 1, 2, and 4, respectively.

We also evaluate our column-wise N:M pruning method on additional CNN architectures: ResNet-18, ResNet-34, ResNet-101, ResNet-152, MobileNet-V2, and DenseNet-121. These tests use batch size 1, reflecting typical RISC-V embedded system usage. 
Table ~\ref{tab:model_accuracy_runtime_7subtables} summarizes the inference time and top-1 accuracy for various sparsity ratios.
For ResNet models with fewer than 50 layers, our pruning method achieves up to a 4.0$\times$ speedup over the dense NHWC baseline, with less than 2.1\% accuracy loss at high sparsity ratios and even improve 0.3\% accuracy at 25\% sparsity.
For deeper ResNet models with more than 50 layers, we observe speedups of up to 3.2$\times$ over the dense baseline, with accuracy degradation kept below 1.9\%.

For MobileNet-V2, our pruning method achieves up to a $1.4\times$ speedup over the dense baseline.
However, applying the proposed pruning pattern to MobileNet's depthwise separable architecture results in a drop in top-1 accuracy, with 12\% degradation at 75\% sparsity.
This arises from MobileNet's relatively fewer number of model parameters compared to the other models, thus more sensitive to the structured sparsity than larger CNN architectures.
For DenseNet-121, our pruning method shows modest speedup but maintains accuracy well, with a 0.6\% increase at 50\% sparsity and only a 0.9\% drop at 75\%.
In summary, these results validate the effectiveness of our column-wise N:M pruning approach for accelerating inference with minimal impact on model accuracy across a range of CNN architectures.

\subsection{Performance of Different Data Layout}\label{subsec:expr_data_layout}
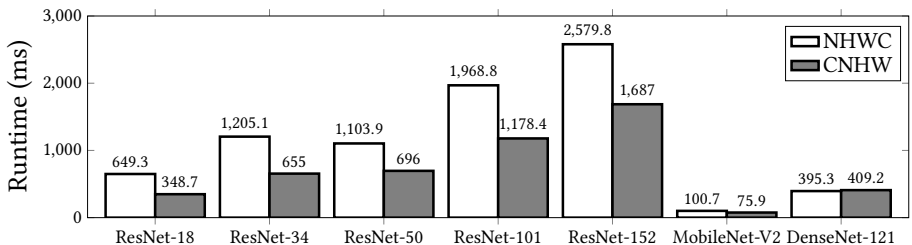
\begin{figure*}[t!]
\centering
\begin{tikzpicture}[scale=0.6]
\begin{axis}[
    width=19.8cm,
    height=6cm,
    ymin=0,
    ymax=3000,
    enlarge y limits=false,
    ybar=0cm,
    bar width=1.1cm,
    enlarge x limits={abs=1.5cm},
    xtick=data,
    tick label style={font=\Large},
    symbolic x coords={
          ResNet-18, ResNet-34, 
          ResNet-50, ResNet-101,
          ResNet-152, MobileNet-V2,
          DenseNet-121,
      },
      nodes near coords={
          \pgfmathprintnumber[fixed,precision=1]{\pgfplotspointmeta}
      },
      every node near coord/.append style={font=\large},
      ylabel={\parbox{5cm}{\centering \huge Runtime (ms)}},
      legend style={at={(0.92,0.95)}, anchor=north, font=\LARGE},
    ]
      \addplot [draw=black, fill=white, ultra thick, area legend] 
        coordinates {
          (ResNet-18,649.300625)
          (ResNet-34,1205.1325)
          (ResNet-50,1103.879286)
          (ResNet-101,1968.785)
          (ResNet-152,2579.75)
          (MobileNet-V2,100.665)
          (DenseNet-121,395.279)
        };
      \addplot [draw=black, fill=gray, ultra thick, area legend] 
        coordinates {
          (ResNet-18,348.6965714)
          (ResNet-34,655.0372222)
          (ResNet-50,696.007)
          (ResNet-101,1178.432857)
          (ResNet-152,1686.956667)
          (MobileNet-V2,75.934)
          (DenseNet-121,409.1957143)
        };
\legend{NHWC, CNHW}
\end{axis}
\end{tikzpicture}
\vspace{-7mm}
\caption{Comparison of inference time with NHWC and CNHW layouts.}
\label{fig:model_layout_runtime}
\Description{Runtime comparison of runtimes(ms) across NHWC and mixed NHWC/CNHW layouts for different models.}
\vspace{-5mm}
\end{figure*}

This section compares the end-to-end performance of dense convolutional networks using the standard NHWC layout (as implemented in the SiFive-optimized XNNPACK) versus the proposed CNHW layout.  Experiments were conducted on ResNet-18, ResNet-34, ResNet-50, ResNet-101, ResNet-152, MobileNet-V2, and DenseNet-121, with the vector length multiplier fixed at LMUL = 4.

Figure~\ref{fig:model_layout_runtime} summarizes the results. For ResNets with fewer than 50 layers, the CNHW layout yields speedups of up to 1.8$\times$ over the NHWC baseline.  For deeper ResNets (with 50 or more layers), the benefit slightly decreases but still reaches up to 1.6$\times$. MobileNet-V2, which consists mostly of lightweight pointwise and depthwise convolutions, sees  a more modest acceleration of approximately 1.3$\times$. In contrast, DenseNet-121 exhibits no measurable improvement and even incurs a minor slowdown.

The observed trends are attributable to differences in memory movement cost. In shallow ResNets, all convolutions are 3x3, so our fused im2col and data packing optimization substantially reduces memory traffic, maximizing the advantage of CNHW.
Deeper ResNets insert two 1x1 convolutions around every 3x3 convolution,  lowering the fraction of compute that can benefit from fusion and thus reducing the speedup.
MobileNet-V2 relies even more heavily on 1x1 convolutions, further limiting the gain.
DenseNet-121 uses small and compact convolutional filters.
In the NHWC baseline, the im2col transformation is applied to weight tensors, whereas in CNHW it is applied to input feature maps.
When the weight tensor is smaller than input feature maps, NHWC entails less data movement, explaining the mild performance drop observed for DenseNet-121.

In summary, the CNHW layout offers the greatest performance benefits for networks dominated by heavy-weight 3x3 convolutions. Its advantage diminishes when the network includes a high proportion of 1x1 convolutions, or when the convolutional weights are significantly smaller than the associated input feature maps. 
\section{Discussion}\label{sec:discussion}

\paragraph{Why we adopt the \textbf{CNHW} layout.}
We choose the \textbf{CNHW} tensor layout for convolution layers because the elements along the $W$ dimension are stored contiguously, enabling efficient vectorized \textit{im2col}.
While the NCHW layout (as used in \cite{elsen2020fast}) is a viable alternative, CNHW offers distinct advantages especially when processing multiple images concurrently (batch size > 1):
\begin{enumerate}[itemsep=0pt,nosep]
  \item \textbf{Efficient conversion from NHWC.}  
        Transformation from the NHWC format to CNHW (and back) involves only two transpose operations. In contrast, converting to NCHW requires additional permutation, incurring extra latency and memory traffic.
  \item \textbf{Improved batch-level packing.}
        In CNHW, data packing can access values from {\em different} batches within a matrix row (input channel), maximizing vector register utilization. NCHW, however, confines each matrix row to a single batch, leading to underutilized vector lanes, particularly when the batch size is small.
\end{enumerate}
These characteristics allow CNHW to better leverage vector hardware, resulting in higher throughput than NCHW.

\section{Conclusion}\label{sec:conclusion}

Weight pruning is a widely adopted technique for reducing the computational demands of deep neural networks.
However, the additional memory accesses introduced by pruning can lead to increased memory overhead, potentially limiting performance gains.
In this work, we introduce column-wise N:M pruning, which groups weights along the weight matrix columns as pruning units.
This approach preserves spatial locality, enhances weight reuse, and scales efficiently to arbitrarily N:M sparsity patterns. 
To further reduce memory movement, we fuse the two most bandwidth-intensive operations, im2col and data packing, into a single pass, thereby reducing memory overhead. 
Our software stack integrates these optimizations into XNNPACK as an extended backend of the AITemplate AI compiler, leveraging AITemplate’s auto-tuning to select the optimal tile sizes and vector register group multipliers for efficient matrix multiplication.
We demonstrate the potential of N:M pruning combined with layout and compiler optimizations to significantly accelerate inference on RISC-V-based platforms without compromising accuracy.
\bibliographystyle{plainnat}
\bibliography{reference}

\end{document}